\documentclass[preprint,showpacs,12pt]{revtex4}
\usepackage{amsmath,amssymb,amsfonts}
\usepackage{graphicx}
\usepackage{hhline}
\usepackage{bm}
\usepackage{amsmath}
\usepackage{epsfig}
\usepackage{epstopdf}

\begin{document}
\begin{flushleft}
\end{flushleft}
\title{Gamow-Teller strength in deformed nuclei within the self-consistent charge-exchange quasi-particle random-phase approximation with the Gogny force}
\author {M. Martini}
\affiliation{Institut d'Astronomie et d'Astrophysique, CP-226, Universit\'e Libre de Bruxelles, 1050 Brussels, Belgium}
\affiliation{Department of Physics and Astronomy, Ghent University, Proeftuinstraat 86, B-9000 Gent, Belgium}
\affiliation{CEA/DAM/DIF, F-91297 Arpajon, France}
\author {S. P\'eru}
\affiliation{CEA/DAM/DIF, F-91297 Arpajon, France}
\author {S. Goriely}
\affiliation{Institut d'Astronomie et d'Astrophysique, CP-226, Universit\'e Libre de Bruxelles, 1050 Brussels, Belgium}

\begin{abstract}
The charge-exchange excitations in nuclei are studied within the fully self-consistent proton-neutron quasiparticle random-phase approximation using the finite-range Gogny interaction. 
No additional parameters beyond those included in the effective nuclear force are included.
Axially symmetric deformations are consistently taken into account, both in the description of the ground states and spin-isospin excitations. We focus on the isobaric analog and Gamow-Teller resonances. A comparison of the predicted strength distributions to the existing experimental data is presented and the role of nuclear deformation analyzed. 
The Gamow-Teller strength is used to estimate the $\beta^-$-decay half-life of nuclei for which experimental data exist. A satisfactory agreement with experimental half-lives is found and justifies the additional study of the exotic neutron-rich $N=82$, 126 and 184 isotonic chains of relevance for the r-process nucleosynthesis.
\end{abstract}

\maketitle


\section{Introduction}     

Spin-isospin nuclear excitations \cite{Osterfeld:1991ii}, in particular the Gamow-Teller (GT) resonances, nowadays play a crucial role in several fields of physics.
First in fundamental  nuclear physics by providing information on the nuclear interaction, 
the equation of state of asymmetric nuclear matter as well as the nuclear skin thickness \cite{Paar:2007bk}. 
Second, in astrophysics where they govern $\beta$-decay, electron and neutrino capture processes, hence stellar evolution and nucleosynthesis \cite{arnould07,Langanke:2002ab}. 
Finally, in particle physics in connection with the evaluation of the $V_{ud}$ element and the unitarity of the Cabibbo-Kobayashi-Maskawa quark-mixing matrix 
\cite{Towner:2010zz}, on the one hand, and with neutrino physics beyond the standard model 
(neutrinoless double beta decay \cite{Suhonen:1998ck,Avignone:2007fu,Bilenky:2012qi} and neutrino oscillation \cite{Frekers:2011zz,Giunti:2012tn}), on the other hand. 

Experimentally the spin-isospin nuclear excitations are studied via charge-exchange 
reactions, such as (p,n), (n,p), (d,$^2$He), ($^3$He,t) or (t,$^3$He) and  $\beta$-decay measurements. 
In spite of the great efforts and interest, the whole nuclear chart is still not experimentally accessible, so that for the exotic nuclei, one can rely on theoretical models only. 
In this context one of the most popular models is the so-called proton-neutron quasiparticle random-phase approximation (pnQRPA), first introduced in Ref.~\cite{Halbleib}. For a reliable prediction of the spin-isospin nuclear excitations, especially for experimentally unknown nuclei,  two main features of the theoretical model are in order: 
the possibility to deal with deformed nuclei and the use of a unique effective nuclear force. The term unique has  two different meanings here. 
First of all it implies that the interaction is the same for all nuclei,
second that the nuclear interaction used to describe the ground and excited states is the same;
this latter property is usually referred as the self-consistency of the calculation. 
Despite the relatively large number of pnQRPA calculations (see, e.g., Refs. \cite{Moeller,Hirsch:1990yj,Hirsch,Raduta:1991ey,Toivanen:1995zi,Sarriguren,Bender:2001up,Vretenar:2003gn,Paar:2004re,Fracasso:2005yi,Liang:2008tv,Minato:2013oua,Yoshida2013,Mustonen2013} and references therein), 
the number of models nowadays including both features remains small. 
Furthermore, even in the limited number of self-consistent calculations
performed either with the zero-range Skyrme-type forces or in the relativistic mean field framework,
there often remains a coupling constant, 
typically in the particle-particle channel, which is treated as a free parameter usually adjusted to $\beta$-decay half-lives or to the position of GT excitation energies.
The possibility to take into account the nuclear deformation is also very important. 
The $\beta$-decay properties of exotic neutron-rich nuclei (in particular those of interest to the r-process nucleosynthesis \cite{arnould07}) as well as the nuclear matrix elements for the double $\beta$-decay have been shown to  depend significantly on the deformation parameter \cite{Sarriguren:2010bi,Sarriguren:2012jm,Yoshida2013,Mustonen2013}. 
Furthermore, deformed nuclei present a strong fragmentation in the response functions and different nuclear shapes can be experimentally distinguished. 

Here, we present a fully self-consistent axially-symmetric-deformed pnQRPA calculation 
without any additional parameters beyond those characterizing the effective nuclear force, namely the finite-range Gogny force within its two parametrizations, D1M \cite{Goriely:2009zz} and D1S \cite{Decharge:1979fa}. 
This work represents a transposition to the charge-exchange sector of the fully consistent axially-symmetric-deformed QRPA calculations with the Gogny force, first presented in Ref. \cite{Peru:2008gd} and devoted to the study of electromagnetic excitations in deformed nuclei \cite{Peru:2011zz,Martini:2011gy}. 
In Sec.~\ref{sect_form}, the pnQRPA formalism is detailed. In Sec.~\ref{sect_res}, the resulting GT and  isobaric analog resonance (IAR) strength are analyzed and compared to the experimental data. Based on the GT strength, the $\beta^-$-decay half-lives are predicted and compared to the experimental data and other models in Sec.~\ref{sect_t12}. Finally, conclusions and perspectives are given in Sec.~\ref{sect_conc}.

\section{Formalism}
\label{sect_form}

Our approach is based on the pnQRPA on top of axially-symmetric-deformed Hartree-Fock-Bogoliubov (HFB) calculations. 
The HFB equations are solved in a finite harmonic oscillator basis.
As a consequence, the positive energy continuum is discretized. 
All HFB quasiparticle states are used to generate the two-quasiparticle (2-qp) excitations.
This means that in principle our calculation can be performed without a cut in energy or in occupation probabilities. 
According to the symmetries imposed in the present axially-symmetric-deformed HFB calculations in even-even nuclei,
the projection $K$ of the angular momentum $J$ on the  symmetry axis and the parity $\Pi$ are good quantum numbers.
Consequently, pnQRPA calculations can be performed separately in each $K^{\Pi}$ block. 
In this context, phonons are characterized by the excitation operator
\begin{equation}\label{thetaplus}
\theta^+_{\alpha,K}=\sum_{pn} X^{pn}_{\alpha,K} \eta^+_p \eta^+_n-(-)^K Y^{pn}_{\alpha,K} \eta_n \eta_p,
\end{equation}
where $\eta^+$ and $\eta$ are the quasi-particle operators, related to the particle creation ($c^+$) and annihilation ($c$) operators through the $u$ and $v$ Bogoliubov transformation matrices; for example, 
\begin{equation}\label{bogo_transf}
\eta^+_p=u_{p \pi} c^+_\pi-v_{p \pi} c_\pi.
\end{equation}
Here and in the following, repeated indices are implicitly summed over; $p$, $n$ and $\pi$, $\nu$ denote proton and neutron quasiparticle 
and harmonic oscillator states, respectively. 
In the well-known QRPA matrix equation
\begin{equation}\label{equaref}
\left(\begin{array}{cc} { A}& { B}\\{ B}&{ A} \end{array}  \right)
 \left(\begin{array}{c}{X_{\alpha,K}}\\ {Y_{\alpha,K}}\end{array}\right)
= \omega_{\alpha,K} \left(\begin{array}{c}X_{\alpha,K}\\-Y_{\alpha,K}\end{array}\right),
\end{equation}
where  $\omega_{\alpha,K}$ are the energies of the pnQRPA excited  states of the parent nucleus, 
the matrices $A$ and $B$ take, in the case of charge-exchange excitations, the following form:
\begin{eqnarray}
A_{pn,p'n'}&=&(\epsilon_p+\epsilon_n)\delta_{pp'}\delta_{nn'}\nonumber\\
&+&u_{p\pi}v_{n\nu}u_{p'\pi'}v_{n'\nu'}(\langle \pi \nu'\vert V\vert \nu \pi'\rangle -  \langle \pi \nu'\vert V\vert \pi'\nu\rangle)\nonumber\\
&+&v_{p\pi}u_{n\nu}v_{p'\pi'}u_{n'\nu'}(\langle \nu \pi'\vert V\vert \pi \nu'\rangle -  \langle \nu \pi'\vert V\vert \nu'\pi\rangle)\nonumber\\
&+&u_{p\pi}u_{n\nu}u_{p'\pi'}u_{n'\nu'}(\langle \pi \nu\vert V\vert \pi'\nu'\rangle -  \langle \pi\nu\vert V\vert \nu'\pi'\rangle)\nonumber\\
&+&v_{p\pi}v_{n\nu}v_{p'\pi'}v_{n'\nu'}(\langle \pi'\nu'\vert V\vert \pi\nu\rangle -  \langle \pi'\nu'\vert V\vert \nu\pi\rangle)\nonumber\\
\end{eqnarray}
and
\begin{eqnarray}
B_{pn,p'n'}&=&u_{p\pi}v_{n\nu}v_{p'\pi'}u_{n'\nu'}(\langle \pi\nu'\vert V\vert \nu\pi'\rangle -  \langle \pi\nu'\vert V\vert \pi'\nu\rangle)\nonumber\\
&+&v_{p\pi}u_{n\nu}u_{p'\pi'}v_{n'\nu'}(\langle \nu\pi'\vert V\vert \pi\nu'\rangle -  \langle \nu\pi'\vert V\vert \nu'\pi\rangle)\nonumber\\
&+&u_{p\pi}u_{n\nu}v_{p'\pi'}v_{n'\nu'}(\langle \pi\nu\vert V\vert \nu'\pi'\rangle -  \langle \pi\nu\vert V\vert \pi'\nu'\rangle)\nonumber\\
&+&v_{p\pi}v_{n\nu}u_{p'\pi'}u_{n'\nu'}(\langle \pi'\nu'\vert V\vert \nu\pi\rangle -  \langle \pi'\nu'\vert V\vert \pi\nu\rangle).\nonumber\\
\end{eqnarray}
As already emphasized, we use the same nucleon-nucleon effective Gogny force (more exactly the D1M or D1S parametrizations), both for the HFB and QRPA calculations in all particle-hole (ph),
particle-particle (pp), and hole-hole (hh) channels.
This procedure is important to avoid numerical and physical inconsistencies. 
To solve the QRPA matrix equation we use the same numerical procedure recently applied to study 
the giant resonances of the heavy deformed $^{238}$U \cite{Peru:2011zz}. It is based on a massive parallel master-slave algorithm. For a single solution of Eq. (\ref{equaref}) the QRPA provides the set of amplitudes $X_{\alpha,K}$ and $Y_{\alpha,K}$
describing the wave function of the excited state $\vert \alpha,K \rangle= \theta^+_{\alpha,K} \vert 0 \rangle$ in terms of the 2-qp excitations. 

Once the pnQRPA matrix equation is solved we can calculate the response to the Fermi, or isospin lowering, operator 
\begin{equation}
\hat{O}_{IAR}=\sum_{i=1}^{A}\tau_-(i)
\end{equation}
obtaining the IAR, the simplest charge-exchange transition in which a neutron is changed into a proton without any other variation 
of the quantum numbers. In an axially-symmetric-deformed nuclear system, the response function of  a given  $J^\Pi$
contains different $K^\Pi =0^\Pi,\pm 1^\Pi,...,\pm J^\Pi$ components.
In spherical nuclei, all these components are degenerate in energy, so that
the response functions associated with any multipolarity can be directly deduced from the $K^\Pi=0^{\pm}$ result. 
In the case of the IAR the $J^{\Pi}=0^+$ distribution is obtained performing the pnQRPA calculation for $K^{\Pi}=0^+$. For the GT excitations, the external operator reads
\begin{equation}
\hat{O}_{GT}=\sum_{i=1}^{A}\vec{\sigma}(i)~\tau_-(i) 
\end{equation}
generating a spin-flip $(\Delta S = \Delta J =1)$ response. In this case, the  GT $J^{\Pi}=1^+$ distributions are obtained 
by adding twice the $K^{\Pi}=1^+$ component to the $K^{\Pi}=0^+$ result. 
Details to go from intrinsic to laboratory frame can be found in Ref.~\cite{Peru:2008gd}. 
 
\begin{figure}
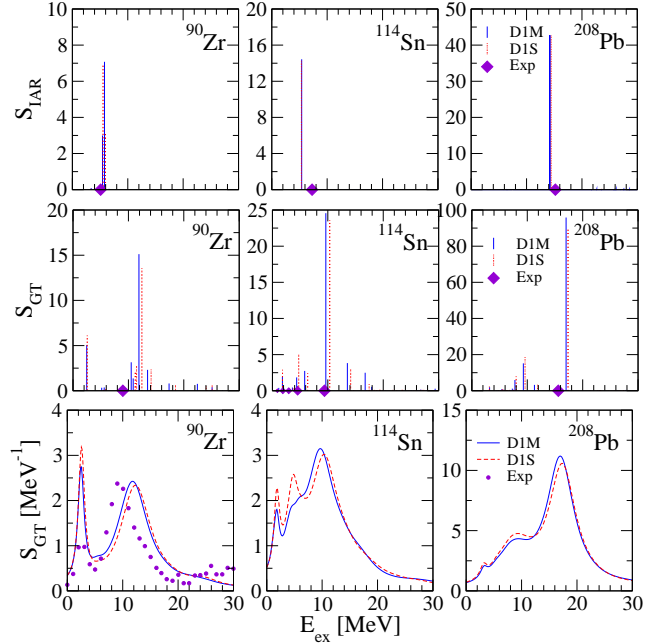

\begin{center}
     \includegraphics[clip,scale=0.35]{fig_zr_sn_pb_iar_new.eps}
     \includegraphics[clip,scale=0.35]{fig_zr_sn_pb_gt_new.eps}
     \includegraphics[clip,scale=0.35]{fig_zr_sn_pb_gt_fold.eps}
\caption{(Color online) pnQRPA  Fermi (upper panels) and GT (middel panels) strength distributions in $^{90}$Zr, $^{114}$Sn and $^{208}$Pb calculated with the 
D1M and D1S forces. The experimental energy peaks obtained from scattering data \cite{Wakasa:1997zz,Akimune:1995zz,Pham:1995zz} are shown as diamonds on the $x$-axis. The lower panels show the folded GT strengths and the comparison with experimental data available for $^{90}$Zr \cite{Wakasa:1997zz}.}
\label{fig_zr_sn_pb}
\end{center}
\end{figure}

\section{Results}
\label{sect_res}
As test case, we first consider the closed neutron-shell nuclei $^{90}$Zr and $^{208}$Pb, as well as  neutron open-shell nucleus $^{114}$Sn. 
In the upper and middle panels of Fig. \ref{fig_zr_sn_pb}, their Fermi and GT strength distributions calculated with D1M and D1S interactions are compared to the 
experimental data \cite{Wakasa:1997zz,Akimune:1995zz,Pham:1995zz}. Even if, in principle, our calculation can be performed without a cut in energy,  in practice we consider here
2-qp states up to an energy of 70 MeV which turns out to be large enough to totally exhaust the Fermi and Ikeda sum rules. 
The results are expressed as a function of the excitation energy $E_{ex}$ referred to the ground state of the daughter nucleus. 
In our model, it is obtained by subtracting a reference energy $E_0$ from the 
excitation energy $\omega_{\alpha,K}$ of the parent nucleus calculated in the pnQRPA, i.e., $E_{ex}=\omega_{\alpha,K}-E_0$. 
The reference energy corresponds to the lowest 2-qp excitation associated with the ground state of the odd-odd daughter nucleus in which the  quantum numbers of the single quasi-proton and neutron states are obtained from the self-consistent HFB calculation of the odd-odd system. 
We remind that for even-even nuclei, the HFB ground state  $\vert \tilde{0}\rangle $ is assumed to be an independent quasi-particle state
$\vert \tilde{0}\rangle =\displaystyle \prod_{i=1} \eta_i\vert - \rangle$ (quasi-particle vacuum).
However, for an odd-odd system, the HFB equations involve a ground state $\vert \pi \nu \rangle$ described as a 2-qp (proton neutron) excitation on top of a qp vacuum $\vert \tilde{0}_{\pi\nu}\rangle$: $ \vert \pi \nu\rangle = \eta'^{+}_\pi \eta'^{+}_\nu  \vert \tilde{0}_{\pi\nu}\rangle $ with $\vert \tilde{0}_{\pi\nu}\rangle = \displaystyle \prod_{i=1} \eta'_i\vert - \rangle $; $\pi$ ($\nu$) running over proton (neutron) qp states. 
In practice, we perform several ``blocked'' HFB calculations (obtained through the minimization of the total binding energy with respect to the ground state $\vert \pi \nu \rangle$), each of them corresponding to a specific choice of the proton and neutron qp quantum numbers.  The couple ($\eta'_\pi$, $\eta'_\nu $) that gives the lowest binding energy among the different HFB calculations is selected, and the corresponding quantum numbers of the odd-odd HFB ground-state (spin and parity) deduced. Such a procedure allows us to determine consistently the quantum characteristics of the reference 2-qp excitation in the parent nucleus. 
In most cases, the reference energy $E_0$ is equal to the lowest energy of the 2-qp excitation of ph type.

Both interactions give  quite similar results for the position of the main peak. 
A one-to-one correspondence between the predicted main peaks is found. 
The energy position of the experimental IAR is quite well reproduced. 
The IAR is experimentally known to be characterized by a single narrow state. This is the case not only of the $^{208}$Pb but also for the open shell $^{114}$Sn. 
The result for $^{114}$Sn reflects the right contribution of the pp channel to the proton-neutron residual interaction, 
without which the response function will be fragmented \cite{Paar:2004re}. The situation is slightly different for  $^{90}$Zr where two states 
very close to each other, probably experimentally undistinguishable, appear. 
Note that our HFB calculations only includes the direct contribution of the Coulomb interaction, while the Coulomb exchange part is not taken into account. This approximation overestimates the proton pairing in general \cite{robledo2001}, and in $^{90}$Zr in particular. 
For this nucleus we repeated the calculation of the IAR starting from HFB calculations 
including direct and exchange Coulomb fields and obtained a disappearance of the fragmentation. However, since no Gogny interaction has yet been derived including the Coulomb contribution to the pairing field, we will restrict ourselves here to the standard HFB calculations as a starting point for our pnQRPA calculation. We have also checked that switching-off the Coulomb interactions in HFB calculations brings the IAR down to zero energy. 

Turning to the GT (middle panel of Fig. \ref{fig_zr_sn_pb}), the D1M interaction is seen to give rise to a strength located at lower energies with respect to the one found with D1S. 
For the nuclei analyzed here, this energy shift rarely exceeds 0.5 MeV. In this context, it should be recalled that the D1M and D1S interactions are characterized by rather different parameters, leading to different nuclear matter properties and Landau parameters. 
The GT energy is known to be sensitive to the single-particle spectrum as well as to the Landau parameter, in particular $g_0^{\prime}$  \cite{Bender:2001up,Borzov03,Niu:2012mi}. More specifically, there is a general tendency for the GT energy to increase with increasing spin-orbit strength parameters $W_{LS}$ and with increasing values of $g_0^{\prime}$, though this latter tendency may be less clear (see Ref.~ \cite{Niu:2012mi} for more details). For the interactions considered here, $W_{LS}$=115.4 MeV fm$^5$ for D1M and 130.0 MeV fm$^5$ for D1S while $g_0^{\prime}=0.71$ for D1M and 0.61 for D1S. Even if the total effect on the energy position of the GT peak is a delicate balance between the effects related to the single-particle spectrum (particularly sensitive to the spin-orbit strength), the residual interaction (strictly related to the Landau parameter $g_0^{\prime}$) as well as the $E_0$ shift, the systematic (small) D1S overestimate of the GT energy with respect to D1M seems to suggest that the $W_{LS}$ parameter plays the major role.

As far as the comparison with experimental data is concerned, the agreement  is seen to be rather satisfactory (Fig. \ref{fig_zr_sn_pb}). 
A small but systematic overestimate of the GT peak is found. 
Particle-vibration coupling \cite{Niu:2012mi,Colo:1994nv} as well as tensor interaction contribution \cite{Bai:2009zzb}, 
both absent in our approach, have been shown to lead to a small shift of the giant GT resonance towards lower energies. 

\begin{figure}
\includegraphics[clip,scale=0.4]{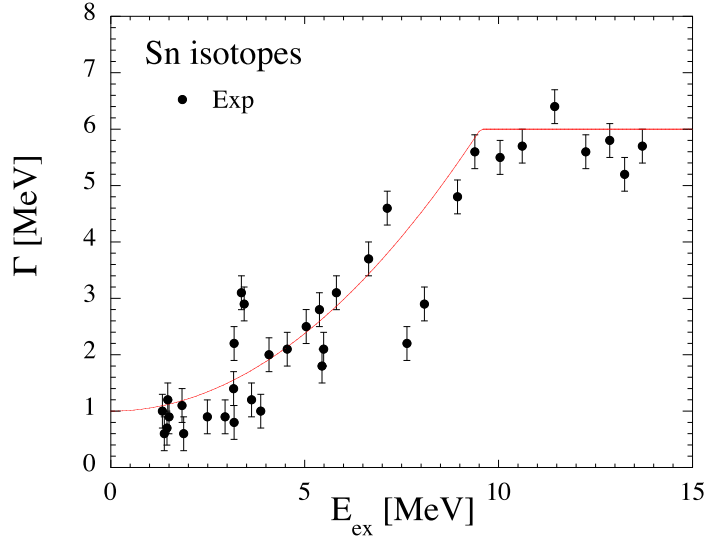}
\caption{\label{fig_Gamma_exp_Sn} (Color online) The experimental GT widths of Sn isotopes \cite{Pham:1995zz} and the adopted parametrization (solid line).}
\end{figure}

Our pnQRPA calculation provide a discrete strength distribution. 
To derive a smooth continuous strength function, the pnQRPA GT strength can be folded with a Lorentz function, as classically done. 
To do so, the spreading width $\Gamma$ is parametrized to reproduce the experimental GT widths found experimentally in Sn isotopes with $A=112-124$ \cite{Pham:1995zz}, as shown in Fig. \ref{fig_Gamma_exp_Sn}. 
The spreading width can be parametrized as $\Gamma[{\rm MeV}]=1+0.055 E_{ex}^2$ (where the excitation energy $E_{ex}$ is expressed in MeV) with an upper value limited to 6~MeV. 

The folded GT strength for $^{90}$Zr, $^{114}$Sn and $^{208}$Pb are shown in the lower panel of Fig. \ref{fig_zr_sn_pb} and compared to the experimental data in the case of $^{90}$Zr \cite{Wakasa:1997zz}. 
The agreement of our calculation with the experiment is reasonable. The double peaks structure, the position of the low energy peak, as well as 
the width of the higher resonance are rather well reproduced while, as already discussed, the centroid energy peak of the higher resonance is overestimated. 
In the case of the  $^{114}$Sn and $^{208}$Pb only experimental counts of the 
($^3$He,t) reaction are available, hence a quantitative comparison of the GT strengths is not straightforward.

\begin{figure}
\includegraphics[clip,scale=0.4]{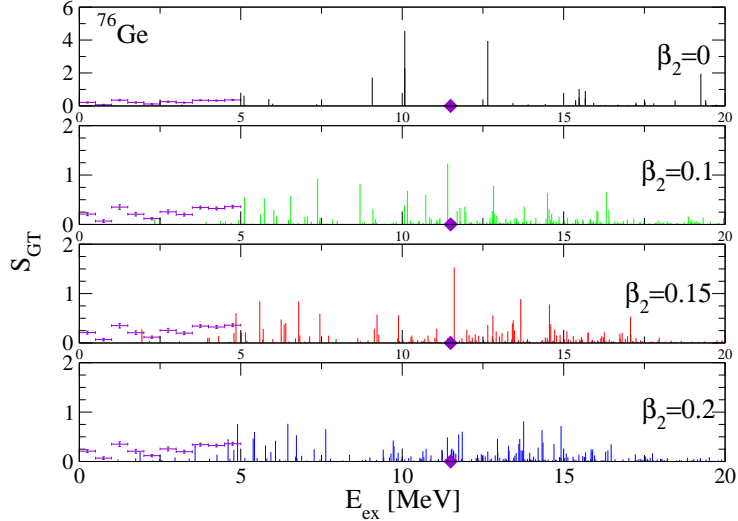}
\caption{(Color online) pnQRPA  GT strength distributions in $^{76}$Ge obtained with the D1M force 
for several values of the deformation parameter $\beta_2$, including the HFB ground state minimum at $\beta_2=0.15$.
The experimental low-energy data \cite{Thies:2012zz} as well as the energy position of the main GT peak are also shown.}
\label{fig_ge76}
\end{figure}

\begin{figure}
\includegraphics[clip,scale=0.4]{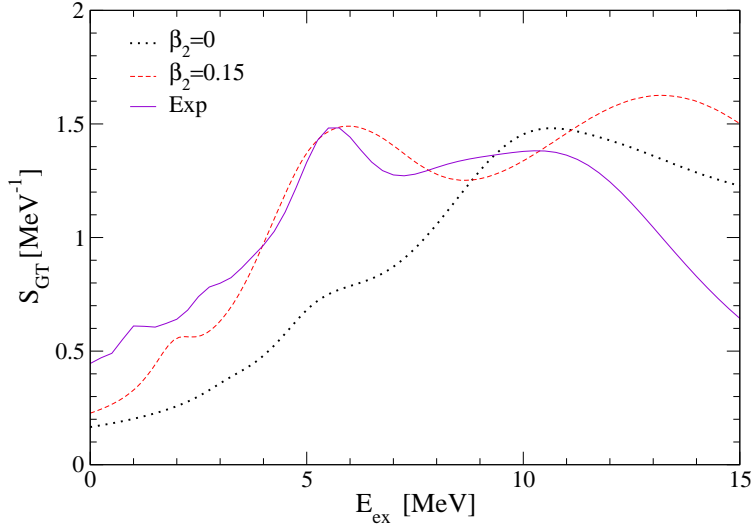}
\caption{(Color online) pnQRPA  GT strength distributions in $^{76}$Ge with the D1M force 
for $\beta_2=0$ and $\beta_2=0.15$ folded as described in the text.
For comparison the experimental strength \cite{Madey:1989zz} folded in a similar way is also given.}
\label{fig_ge76_fold}
\end{figure}

The above results refer to three spherical nuclei.
As already emphasized,  our approach describes axially symmetric deformed nuclei. 
As an example for a deformed nucleus, we consider $^{76}$Ge, a nucleus of particular interest in the neutrinoless double $\beta$-decay experiments in the past \cite{KlapdorKleingrothaus:2000sn,Aalseth:2002rf}, present \cite{Agostini:2013mzu}, and future \cite{Gaitskell:2003zr}.  

We show in Fig.~\ref{fig_ge76} the $^{76}$Ge GT excitations obtained with the D1M interaction for four different values of the quadrupole deformation parameter $\beta_2$, including the HFB minimum at $\beta_2=0.15$. 
As expected, the deformation tends to increase the fragmentation of the response. 
Calculations with different deformations produce peaks that are displaced. 
This is true not only for the giant resonance region but also for the low energy states. 
Recently the low energy part of the GT excitations of the $^{76}$Ge has been studied with high precision \cite{Thies:2012zz} 
due to its importance for the neutrinoless double $\beta$-decay physics. We show this experimental results in Fig.~\ref{fig_ge76} 
to compare to our results at different $\beta_2$. 
It appears that deformation effects influence the low-energy strength and that the spreading of the low-energy GT strength can be rather well reproduced for deformations around $\beta_2=0.10-0.15$, in contrast to what is found in the spherical approximation or at larger deformations. For completeness, we also show in Fig.~\ref{fig_ge76_fold} our folded calculations at $\beta_2=0$ and $\beta_2=0.15$  as well as the experimental results of Ref. \cite{Madey:1989zz} folded in the same way. 
Also in this case the agreement between the experimental data and the $\beta_2=0.15$ case can be considered as satisfactory, at least better than with the spherical case.

\section{Application to half-life calculations}
\label{sect_t12}
\begin{figure*}
\begin{center}
     \includegraphics[clip,scale=0.41]{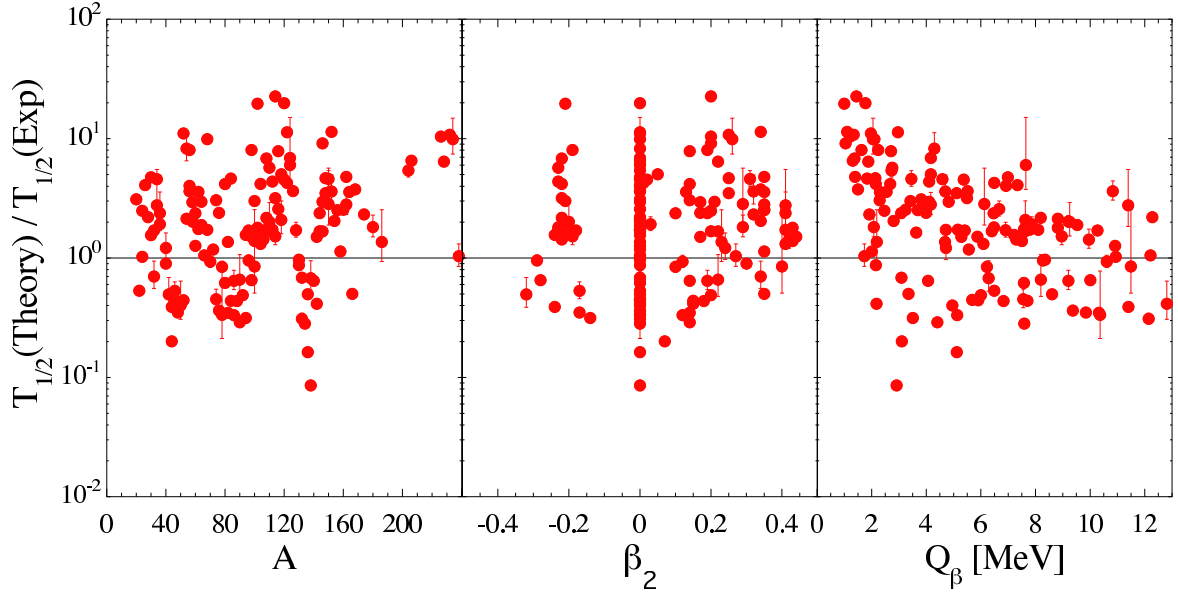}
\caption{(Color online) Ratio between the pnQRPA (obtained with the D1M interaction) and experimental \cite{nubase12} 
$\beta$-decay half-lives as a function of $A$,  $\beta_2$ and $Q_\beta$ for 145 even-even nuclei with an experimental half-life $T_{1/2} \le 1000$~s. Error bars only include experimental uncertainties. }
\label{fig_t12exp}
\end{center}
\end{figure*}
As a first application of our calculation, we now focus on the low-energy GT strength and more specifically on the $\beta^-$-decay half-lives.
In the allowed GT decay approximation the $\beta^-$-decay half-life $T_{1/2}$ can be expressed in terms of the GT strength function $S_{GT}$ according to
\begin{equation}
\frac{\ln 2} {T_{1/2}}=\frac{(g_A/g_V)_{\textrm{eff}}^2}{D} \int_{0}^{Q_\beta}f_0(Z,A,Q_\beta-E_{ex})S_{GT}(E_{ex})dE_{ex}.
\end{equation}
For the phase-space volume $f_0$ as well as  the factor $D$ and  the vector and axial vector coupling constants (including the quenching factor), we refer to the work of Ref.~\cite{Borzov:2000ve}. 
To estimate the $Q_\beta$ mass differences, we take experimental (and recommended) masses \cite{ame12} when available or the D1M mass predictions \cite{Goriely:2009zz}, otherwise.

\begin{figure}
\begin{center}
     \includegraphics[clip,scale=0.4]{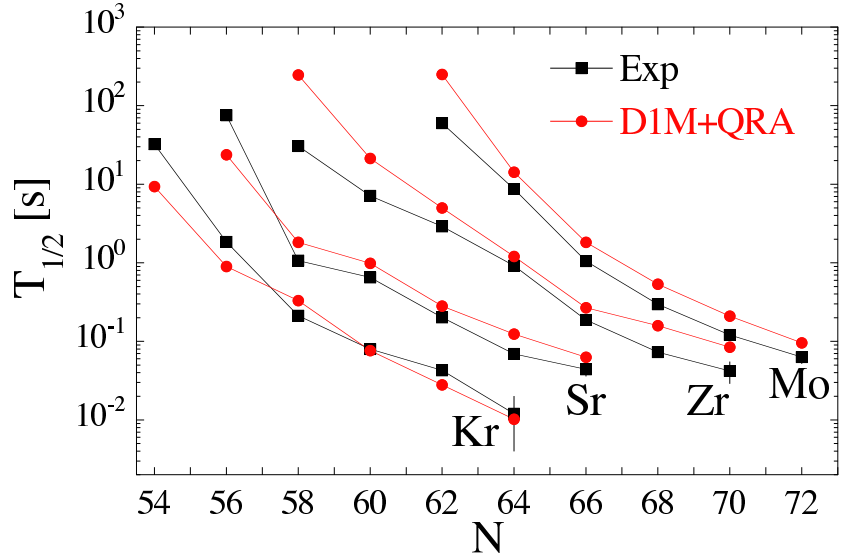}
\caption{(Color online) Comparison between experimental  \cite{nubase12} and D1M+QRPA $\beta$-decay half-life predictions for the known isotopic chains of Kr, Sr, Zr and Mo.}
\label{fig_t12_Kr-Mo}
\end{center}
\end{figure}

To give an idea of the global predictions of our model, we compare in Fig.\ref{fig_t12exp} for even-even nuclei the pnQRPA (obtained with the D1M interaction) $\beta^-$-decay half-lives with the experimental data \cite{nubase12}. 
The results are plotted as a function of the mass number $A$, the deformation parameter $\beta_2$ and the $Q_{\beta}$ value. 
They turn to be quite homogeneous with respect to $A$ and more particularly $\beta_2$.  Larger deviations are found for nuclei close to the valley of $\beta$-stability (Fig.\ref{fig_t12exp}, right panel), i.e for low-$Q_{\beta}$ values, as found in all models.  Note, however, that in Fig.\ref{fig_t12exp} where only nuclei with experimental data are concerned, large $Q_{\beta}$-values essentially correspond to light nuclei for which mean-field models may be less adequate to estimate the ground-state deformation, mixing of configuration being found beyond the mean-field approximation. 
Globally, predictions tend to overestimate the experimental half-lives, but deviations rarely exceed one order of magnitude. 
Note that the half-life overestimation found here is less important that the effect of neglecting pn pairing in relativistic QRPA calculation \cite{Niu:2012rc}. 
We also compare in Fig.~\ref{fig_t12_Kr-Mo} the D1M+QRPA and experimental half-lives for the much studied isotopic chains of Kr, Sr, Zr and Mo which are strongly deformed. Here also, the D1M+QRPA model tends to give rise to half-lives larger than experimental ones, leaving space for possible additional contributions from forbidden transitions.

\begin{figure*}
\begin{center}
     \includegraphics[clip,scale=0.4]{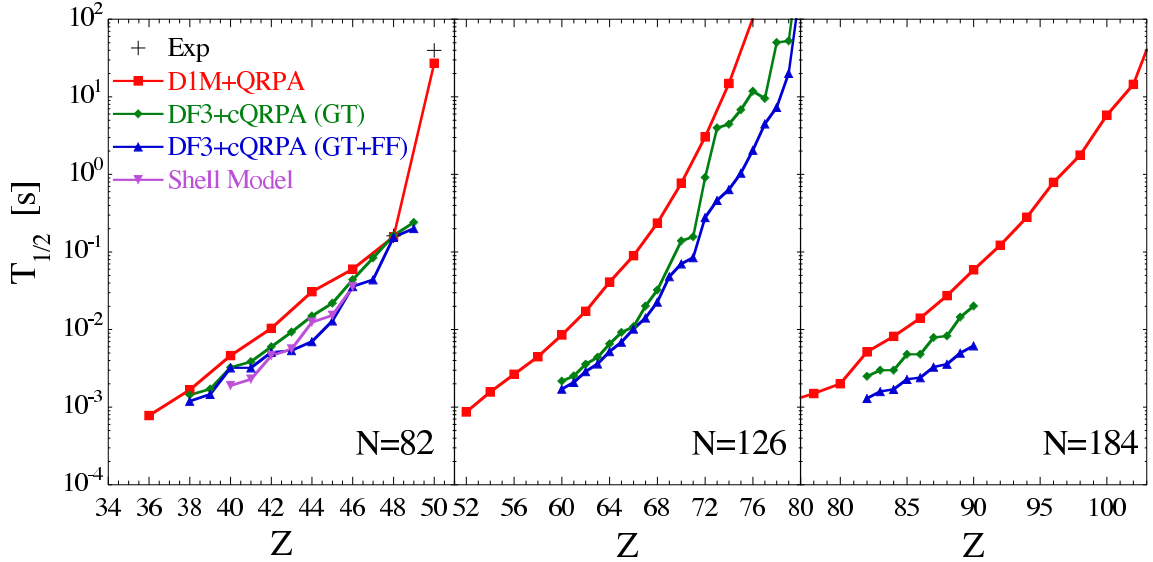}
\caption{(Color online) Comparison between our $\beta$-decay half-life predictions and the DF3+QRPA calculation of \cite{Borzov:2000ve,Borzov:2013}, including the GT contribution or both the GT plus FF contributions, for the neutron-rich nuclei along the $N=82, 126$ and 184 isotones. For the  $N=82$ isotonic chain, experimental data \cite{nubase12} and shell model results \cite{MartinezPinedo:1999rx} are also shown.}
\label{fig_t12_N82}
\end{center}
\end{figure*}

Finally, our $\beta$-decay half-lives are compared to the density function plus continuum QRPA calculation (DF3+cQRPA) of  Ref.~\cite{Borzov:2000ve,Borzov:2013} in Fig. \ref{fig_t12_N82} for the exotic neutron-rich nuclei along the $N=82, 126$ and 184 isotones. We choose to focus on this region of the nuclear chart due to its relevance to the $r-$process nucleosynthesis \cite{arnould07}. Nice agreement with experimental data is found for $^{130}$Cd and $^{132}$Sn. Both the contribution of the GT and the GT plus first-forbidden (FF) transitions are given in Fig. \ref{fig_t12_N82} for the DF3+cQRPA calculation to illustrate the impact of the FF contributions, as predicted by \cite{Borzov:2013}. Clearly, such a contribution need to be included for the $N=184$ nuclei and some of the $N=126$ nuclei.
Our results give rise to decay half-lives systematically larger than the DF3+cQRPA approach. These deviations can originate from different GT strength but also different estimates for the $Q_{\beta}$-values or reference energies $E_0$. We also show in the left panel of Fig. \ref{fig_t12_N82} the shell model predictions \cite{MartinezPinedo:1999rx} for some of the $N=82$ nuclei that are in relatively close agreement with the DF3+cQRPA calculations and lower than ours. Such different predictions could have an impact on the production of the heavy nuclei by the r-process nucleosynthesis, but such an analysis is postponed to a future study. 

\section{Conclusion}
\label{sect_conc}

We present here for the first time a fully self-consistent axially-symmetric-deformed pnQRPA calculation based  on the finite-range Gogny force. 
We applied our model to the analysis of charge-exchange modes paying a special attention to the GT resonances. 
The crucial role of deformation, automatically included in our approach, was analyzed. The agreement with experiment is satisfactory both for the strength functions and the 
$\beta^-$-decay half-lives. Our extrapolation of the $\beta$-decay half-lives to the neutron-rich $N=82$, 126 and 184 isotones of astrophysical interest are found to give rise to larger values with respect to the continuum QRPA calculation of \cite{Borzov:2013}. 
These encouraging results open the way to further studies in several sectors. In particular, it will become possible to include the study of IAR and GT resonances in the procedure of construction and validation of new Gogny-type forces. 
In connection with astrophysics the next step of our work will include forbidden transitions and deal with large-scale calculations of $\beta$-decay half-lives and electron neutrino capture rates  for both even and odd numbers of nucleons to analyze  their impact on the r-process nucleosynthesis. 
Finally, from a particle physics point of view, the evaluation within our model, among others, of the low-energy neutrino-nucleus cross section and  the double-$\beta$ decay nuclear matrix elements  could shed light on the systematics of nuclear origin to be taken into account in these rare processes. 

\begin{acknowledgements}
We thank S. Hilaire for his help in the large-scale computations and I. Borzov for sharing his latest DF3+cQRPA results.
This work was partially supported  by the Communaut\'e Fran\c caise de Belgique (Actions de Recherche Concert\'ees) and the Interuniversity Attraction Poles Programme initiated by the Belgian Science Policy Office (BriX network P7/12). S.G. acknowledges the financial support of the F.N.R.S. 
\end{acknowledgements}

\end{document}